\documentclass[10pt]{article}
\usepackage{graphicx}

\def\Title#1{\begin{center} {\Large #1 } \end{center}}
\def\Author#1{\begin{center}{ \sc #1} \end{center}}
\def\Address#1{\begin{center}{ \it #1} \end{center}}

\newcommand\pubblock{\rightline{\begin{tabular}{l} Proceedings of the Fifth Annual LHCP\\ \pubnumber\\
         \pubdate  \end{tabular}}}

\newenvironment{Abstract}{\begin{quotation} \begin{center} 
             \large ABSTRACT \end{center}\bigskip 
      \begin{center}\begin{large}}{\end{large}\end{center} \end{quotation}}

\newenvironment{Presented}{\begin{quotation} \begin{center} 
             PRESENTED AT\end{center}\bigskip 
      \begin{center}\begin{large}}{\end{large}\end{center} \end{quotation}}

\def\beq{\begin{equation}}
\def\eeq#1{\label{#1}\end{equation}}
\def\eeqn{\end{equation}}

\def\beqa{\begin{eqnarray}}
\def\eeqa#1{\label{#1}\end{eqnarray}}
\def\eeqan{\end{eqnarray}}

\let\bar=\overbar

\def\ie{{\it i.e.}}

\def\Dslash{\not{\hbox{\kern-4pt $D$}}}
\def\dslash{\not{\hbox{\kern-2pt $\del$}}}

\def\msb{{\bar{\ssstyle M \kern -1pt S}}}

\usepackage{color}

\newcommand\pubnumber{ }
\newcommand\pubdate{\today}

\def\affiliation{
Department of Physics and Astronomy\\
University of Southampton, United Kingdom}

\def\support{\footnote{Work supported by the Science and Technology Facilities Council, grant number ST/L000296/1.
All authors acknowledge partial financial support through the NExT Institute.}}

\begin{document}

\large
\begin{titlepage}
\pubblock

\vfill
\Title{Using leptons $p_T$ measurements to constrain $Z^\prime$-bosons widths in Drell-Yan processes at the LHC}
\vfill

\Author{ Juri Fiaschi, Elena Accomando, Stefano Moretti, Claire H. Shepherd-Themistocleous\support }
\Address{\affiliation}
\vfill
\begin{Abstract}

We recognise the appearance of a Focus Point (FP) in the transverse momentum distribution of either leptons originating from a BSM $Z^\prime$-boson decay, after a simple normalisation procedure.
Exploring the properties of the FP we will be able to define in a general way a new observable, the Focus Point Asymmetry ($A_{\rm FP}$), which can be used to set constrains on the  $Z^\prime$ width.
We discuss the potential and the sensitivity of the $A_{\rm FP}$ in as diagnostic tool for $Z^\prime$ physics, considering various $Z^\prime$ phenomenological realisations.

\end{Abstract}
\vfill

\begin{Presented}
The Fifth Annual Conference\\
 on Large Hadron Collider Physics \\
Shanghai Jiao Tong University, Shanghai, China\\ 
May 15-20, 2017
\end{Presented}
\vfill
\end{titlepage}
\def\thefootnote{\fnsymbol{footnote}}
\setcounter{footnote}{0}

\normalsize

\section{Introduction}

The recent upgrade in energy of the LHC is unfolding the high invariant mass region looking for any hint of BSM physics.
At the same time, the steady increase of integrated luminosity is building a very large data set, which requires sharp analysis techniques to 
disentangle new physics signals.

A common prediction of many BSM scenarios is the appearance of an additional massive neutral gauge boson in the spectrum,  often called a $Z^\prime$.
The golden channel for the detection of a $Z^\prime$-boson is the opposite-charge same-flavour two-lepton final state.
Its signal would appear as a peak in the invariant mass distribution of the di-lepton pair, following the shape of a Breit-Wigner (BW) functional
form.
Phenomenologically, $Z^\prime$s can appear in a large variety of realisations.
The most common scenario would be a narrow peak ($\Gamma_{Z^\prime} / M_{Z^\prime} \sim$ 1\%) standing over a null background.
In this picture, the traditional experimental searches based on the ``bump" hunt strategy have maximal sensitivity and from the absence of 
new physics evidences, lower bounds on the $Z^\prime$ mass can be extracted in a model independent way~\cite{Accomando:2013sfa}, such that their 
re-interpretation in other frameworks is unambiguous.
With this approach the ATLAS~\cite{ATLAS:2017wce} and CMS~\cite{CMS:2016abv} collaborations regularly release updated bounds, exploiting the
latest available data.
The most recent results set the mass of the $Z^\prime$-boson to $M_{Z^\prime} >$ 4 TeV for the most of the single $Z^\prime$ benchmark models
(4.5 TeV for the {\rm SSM})~\cite{ATLAS:2017wce}.

The purpose of this article is to introduce a novel observable which can be used to support the analysis of a signal coming from a
BSM $Z^\prime$-boson~\cite{Accomando:2017fmb}.
We will consider the transverse momentum distribution of either lepton in the final state.
We will highlight the appearance of a Focus Point (FP) carrying interesting model independent features and exploiting those properties we will
define a new observable and finally we will examine its discovery and diagnostic potential, especially in the unfriendly scenario
of broad $Z^\prime$ resonances.

\section{The Focus Point and its properties}

As mentioned in the previous section, the current lower bounds on the mass of narrow $Z^\prime$-bosons is around 4 TeV~\cite{ATLAS:2017wce}.
However, those limits apply only if the resonance is narrow, such that the method used to extract those limits breaks down if the resonance width
is enhanced $\Gamma_{Z^\prime} / M_{Z^\prime} >$ 5\%~\cite{Accomando:2017fmb}.
There exist many physical situations where this condition is satisfied. In particular, a mixing in the neutral sector, or the opening
of new exotic decay channels for the $Z^\prime$ resonance, serve this purpose~\cite{Abdallah:2015uba,Accomando:2016sge}.
In essence, the minimal realisations of narrow $Z^\prime$s and their conventional analysis based on the ``bump" hunt, although being a necessary
first step in the quest for detecting new physics signals, usually only covers a small part of the parameter space of a specific model and in general
does not include all the possible phenomenological scenarios.

In the following we will consider $Z^\prime$ signals beyond the narrow width approximation.
In particular, we will refer to some single $Z^\prime$ constructions that are usually adopted as benchmarks models and we will modify the $Z^\prime$
width by hand, which means that the fermions couplings will not be changed (production cross section unchanged), while the partial 
decay branching ratios will be rescaled (following the inverse of the width).
Within this parametrisation, we are going to consider also resonances with masses on the edge of the exclusion limits (or sometimes below), only 
with the purpose of comparing their signal shape with their unexcluded counterparts featuring an enhanced width.
We are also presenting our analysis in the High Luminosity Large Hadron Collider (HL-LHC) setup, as the diagnostic features of the new observable
we are introducing will be exploited in this regime.

\subsection{Profiling $Z^\prime$ signals in the di-lepton channel}

\begin{figure}[h]
\begin{center}
\includegraphics[width=0.45\textwidth]{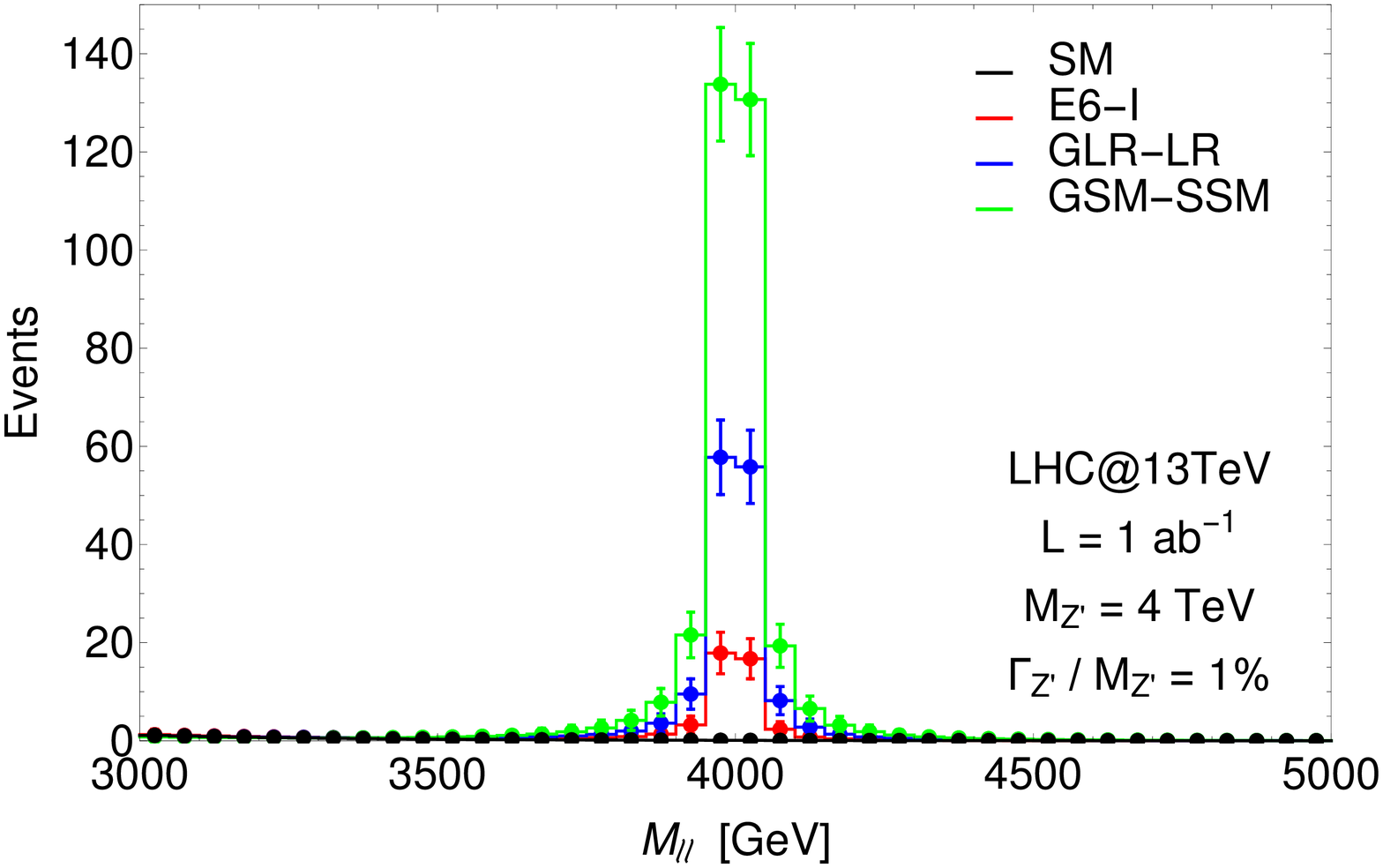}{(a)}
\includegraphics[width=0.45\textwidth]{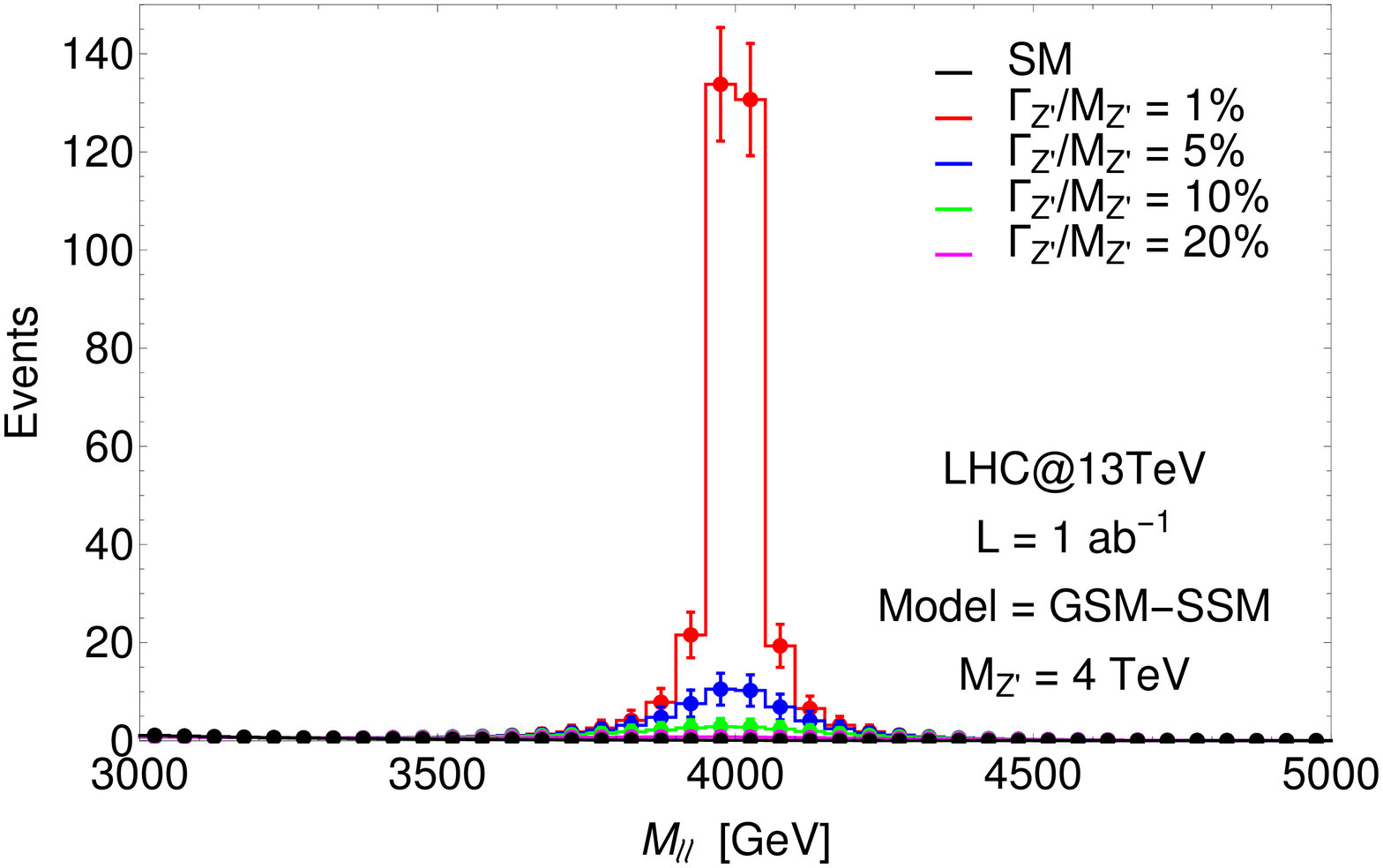}{(b)}
\caption{Distribution of number of events as function of the di-lepton invariant mass as predicted in the SM and 
(a) and in three $Z^\prime$ benchmark models fixing the width of the resonance at 1\% of its mass  
(b) and in the SSM fixing the width of the resonances at four different values (1\%, 5\%, 10\% and 20\% of the mass)
and $M_{Z^\prime}$ = 4 TeV, for the 13 TeV LHC and with $\mathcal{L}$ = 1 ab$^{-1}$. 
Acceptance cuts are applied ($|\eta|<2.5$), detector efficiencies are not accounted for.}
\label{fig:peaks_Minv}
\end{center}
\end{figure}

The profile of a $Z^\prime$ resonance in the di-lepton invariant mass spectrum is visible in Fig.~\ref{fig:peaks_Minv}(a), where three popular
benchmarks have been selected, the {\rm E6-I} model, the Left-Right Symmetric model ({\rm LR}) and the Sequential Standard Model ({\rm SSM}),
fixing the resonance mass at $M_{Z^\prime}$ = 4 TeV and its width at $\Gamma_{Z^\prime} / M_{Z^\prime}$ = 1\%.
We are considering an integrated luminosity of $\mathcal{L}$ = 1 ab$^{-1}$, such that, in this case of narrow resonances, the peaks are very visible
with a moderate statistics.
As the resonance width grows, the sensitivity on the BW peak decreases rapidly.
In Fig.~\ref{fig:peaks_Minv}(b) we are considering one benchmark model (the {\rm SSM}) with the $Z^\prime$ mass fixed 
at $M_{Z^\prime}$ = 4 TeV, while its width over mass ratio varies between 1\% and 20\%.
Already for $\Gamma_{Z^\prime} / M_{Z^\prime} >$ 5\%, the number of events drops sensibly and the ``bump" hunt strategy fails in detecting
the resonance.

\begin{figure}[h]
\begin{center}
\includegraphics[width=0.45\textwidth]{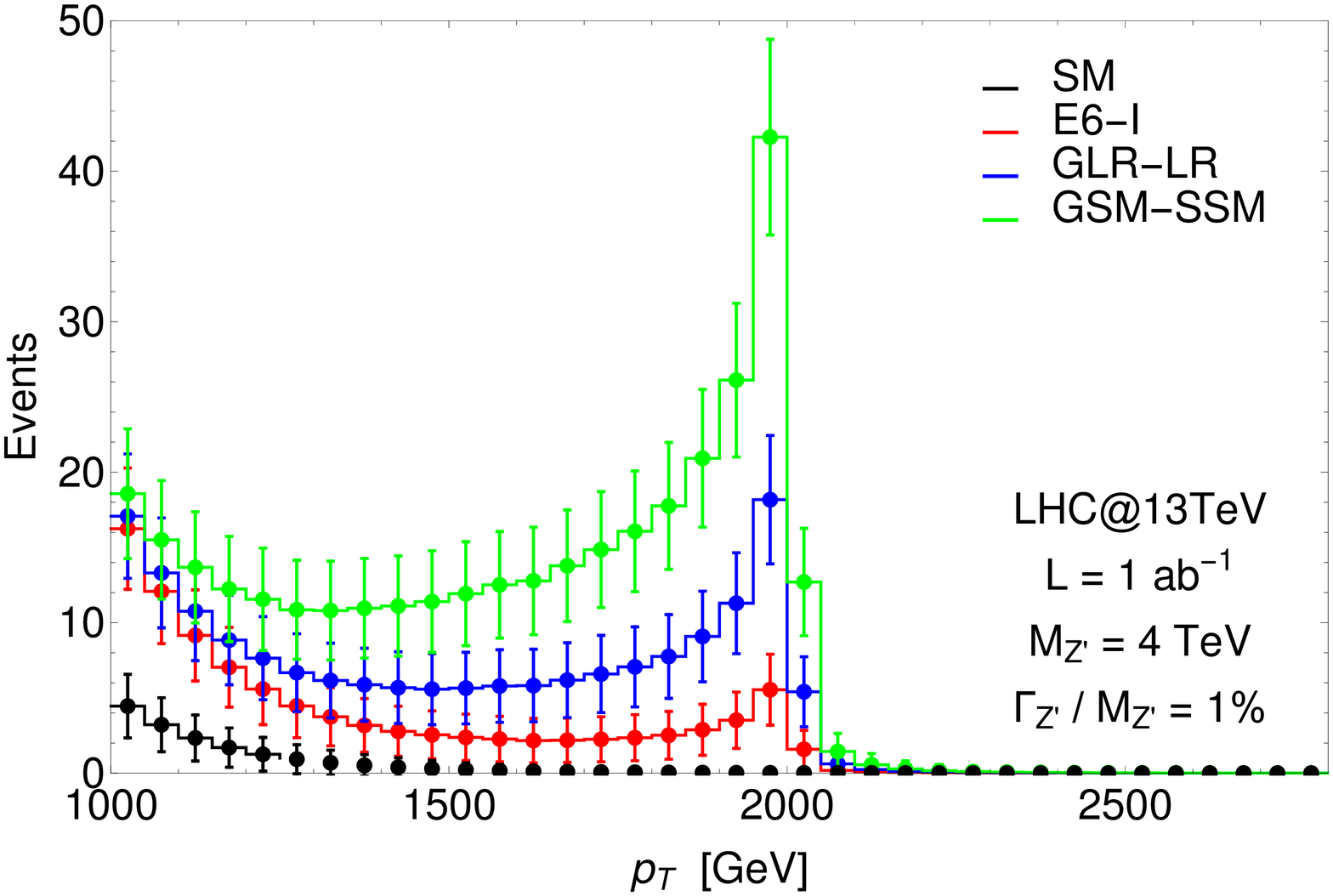}{(a)}
\includegraphics[width=0.45\textwidth]{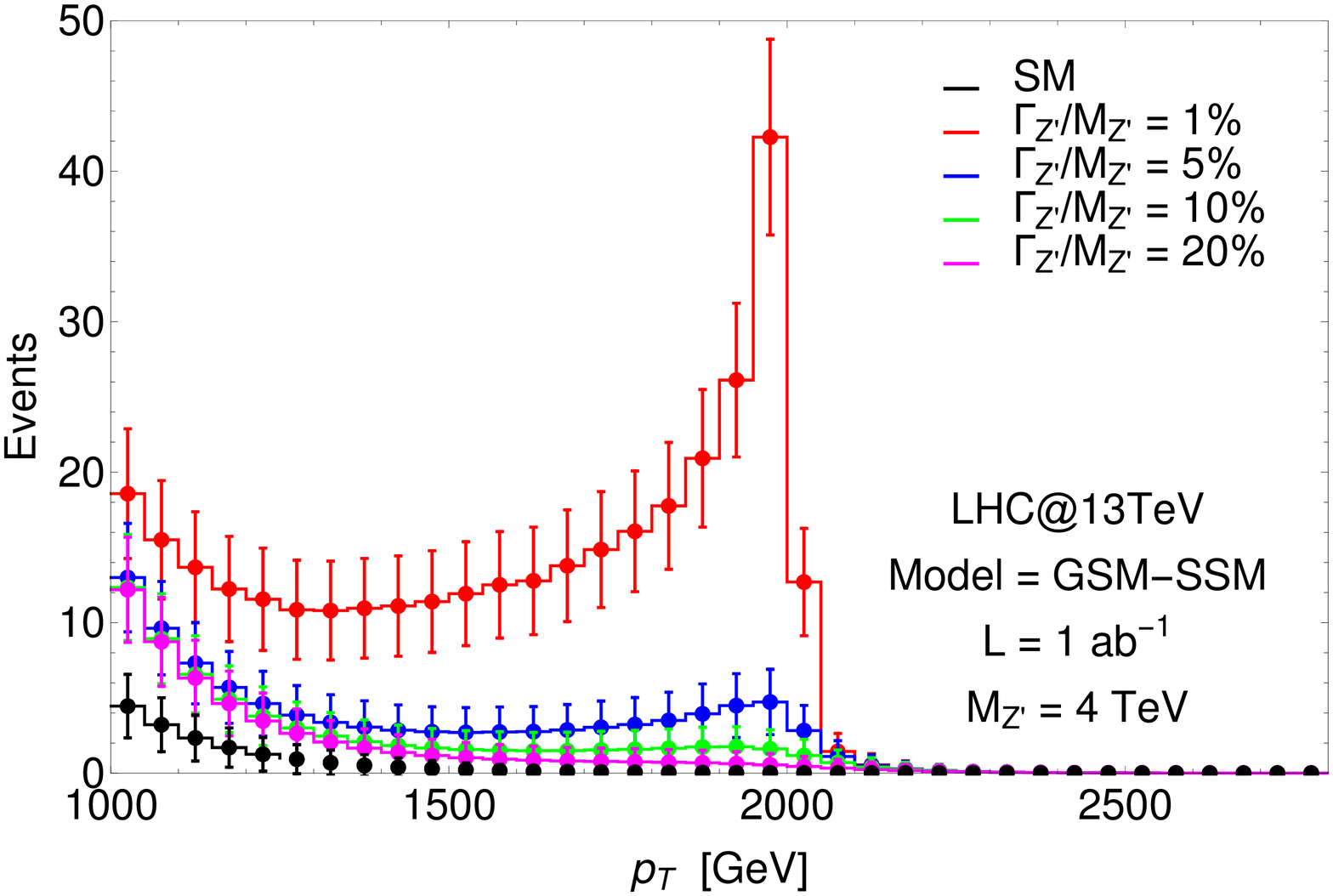}{(b)}
\caption{Same as Fig.~\ref{fig:peaks_Minv} for the distributions of number of events as function of the $p_T$ of either lepton.}
\label{fig:peaks_pT}
\end{center}
\end{figure}

Consider now the transverse momentum distribution of either lepton in the final state.
The $p_T$ distributions of the number of events in the corresponding case of Fig.~\ref{fig:peaks_Minv} are visible in Fig.~\ref{fig:peaks_pT}.
The Jacobian peaks in Fig.~\ref{fig:peaks_pT}(a) are well visible, but observing Fig.~\ref{fig:peaks_pT}(b) it is clear that a shape analysis
of a resonance with $\Gamma_{Z^\prime} / M_{Z^\prime} >$ 5\% would be a difficult task.

\subsection{The Focus Point}

The first step towards the definition of the new observable consists in normalising the curves of Fig.~\ref{fig:peaks_pT}.

\begin{figure}[h]
\begin{center}
\includegraphics[width=0.45\textwidth]{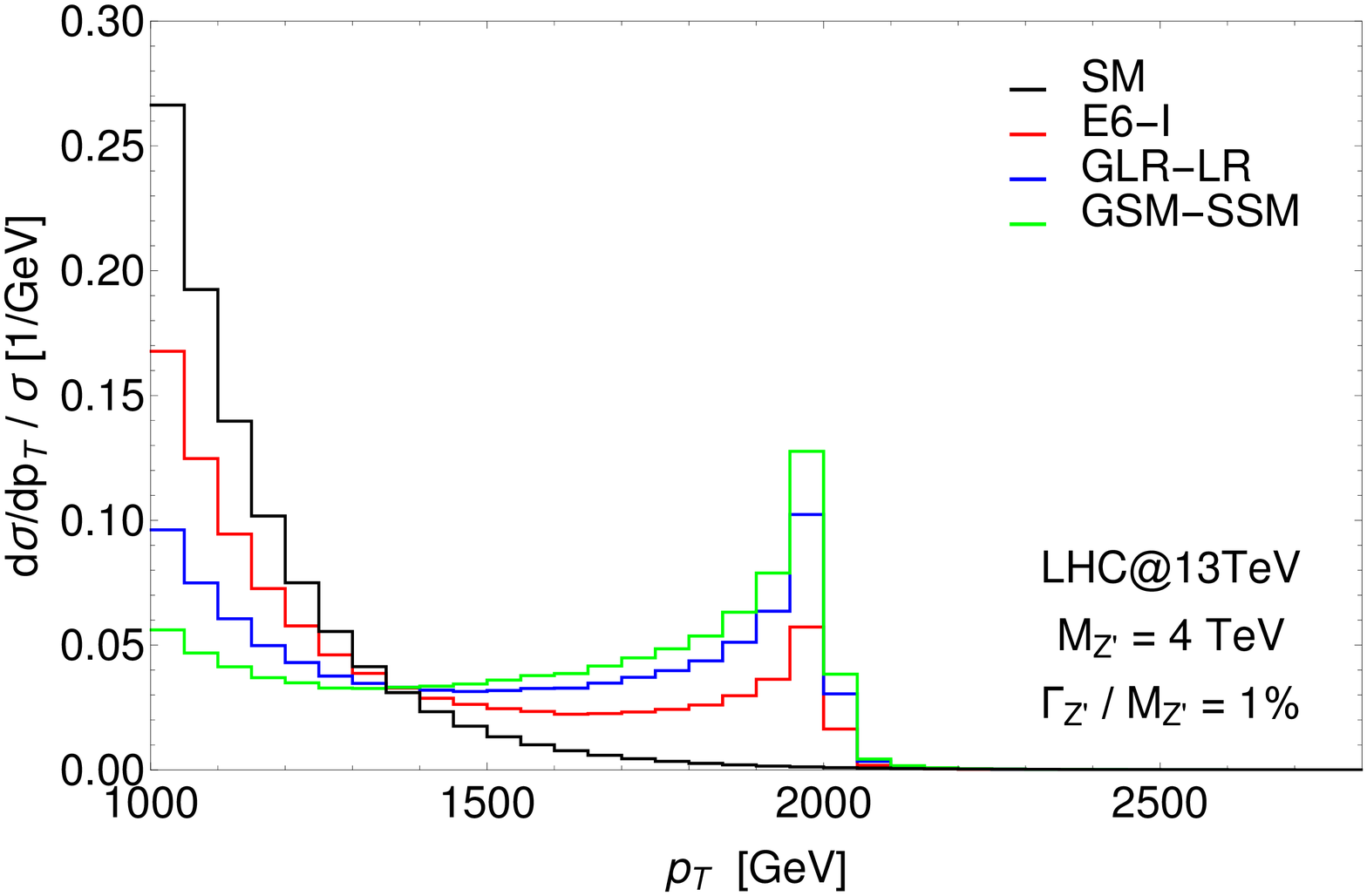}{(a)}
\includegraphics[width=0.45\textwidth]{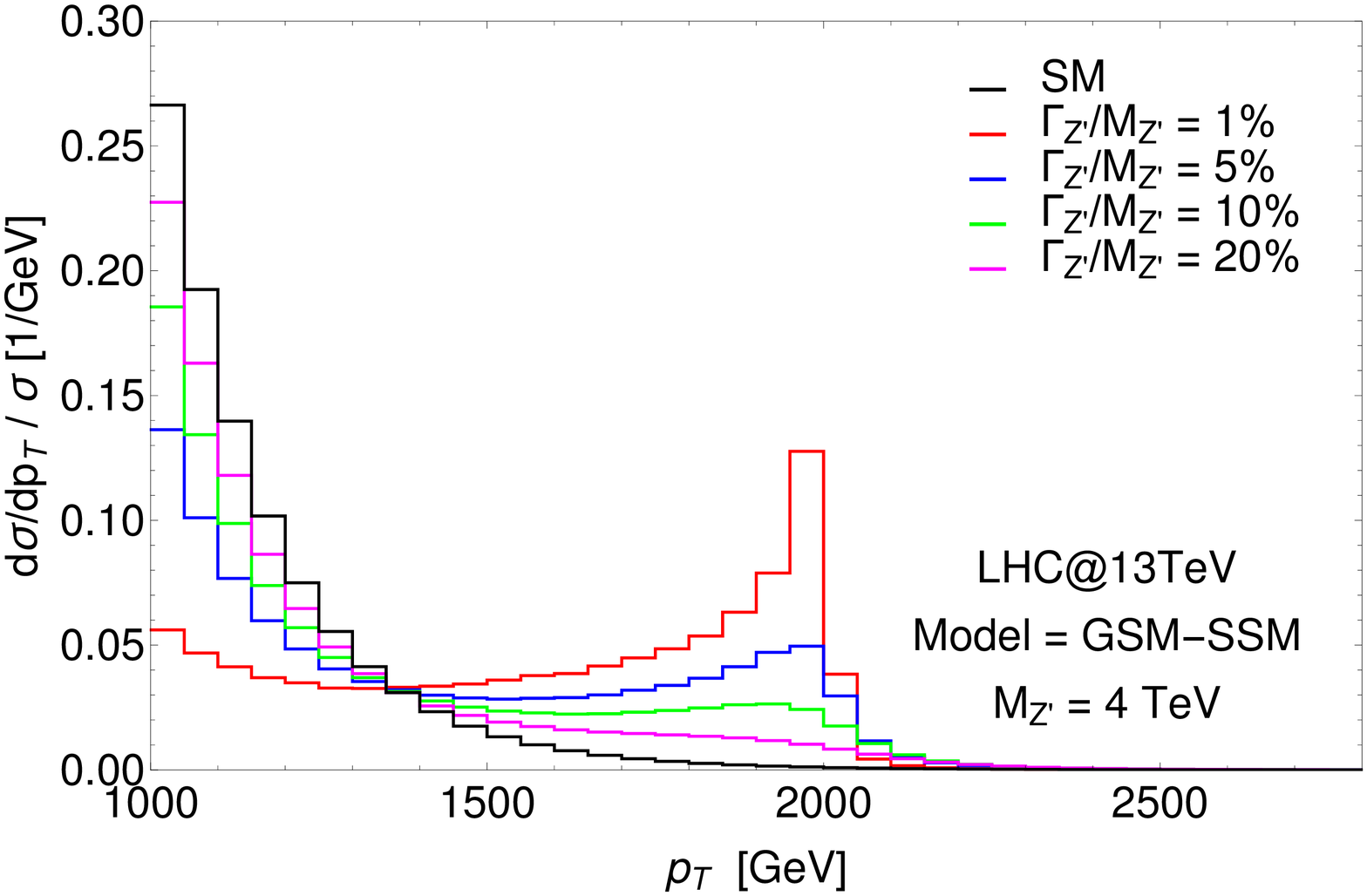}{(b)}
\caption{Normalized distribution obtained from Fig.~\ref{fig:peaks_pT} with $p_T^{\rm min}$ = 1000 GeV.}
\label{fig:pT_norm}
\end{center}
\end{figure}

The result is visible in Fig.~\ref{fig:pT_norm} where the normalisation interval of each curves has been fixed between a $p_T^{\rm min}$ = 1000 GeV,
and a $p_T^{\rm max}$ that can be chosen at any value sufficiently far from the Jacobian peak at $M_{Z^\prime}/2$ to consider all the available data.
The interesting feature appearing from Fig.~\ref{fig:pT_norm} is that now all the curves cross through the same point, that we will call Focus Point (FP).

The FP shows remarkable qualities, already visible in these plots.
In Fig.~\ref{fig:pT_norm}(a) we can see that its position is model independent while from Fig.~\ref{fig:pT_norm}(b) we can recognise that is also 
independent of the resonance width. It is also to notice that the SM background as well crosses the FP.

\begin{figure}[h]
\begin{center}
\includegraphics[width=0.30\textwidth]{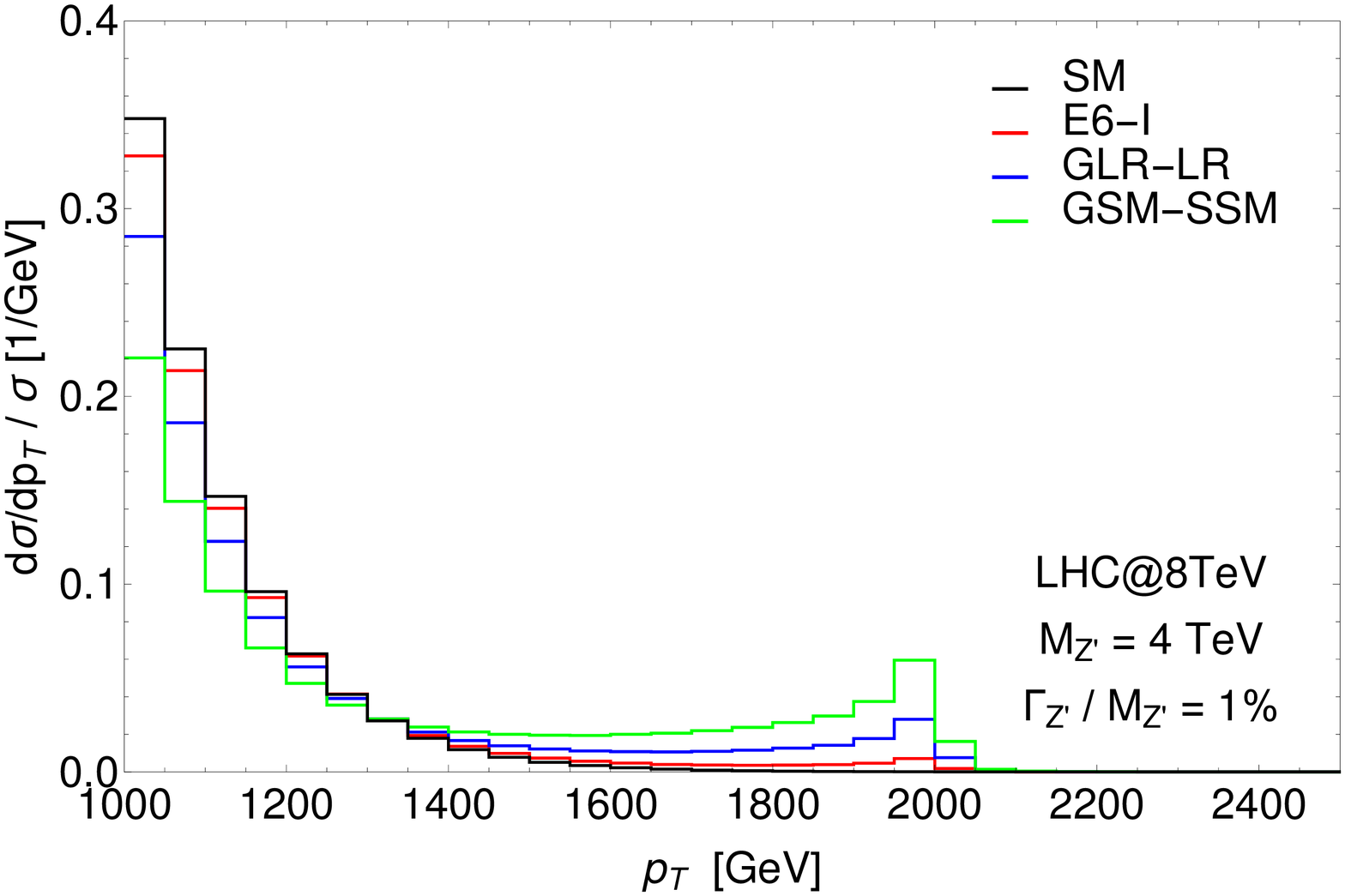}{(a)}
\includegraphics[width=0.30\textwidth]{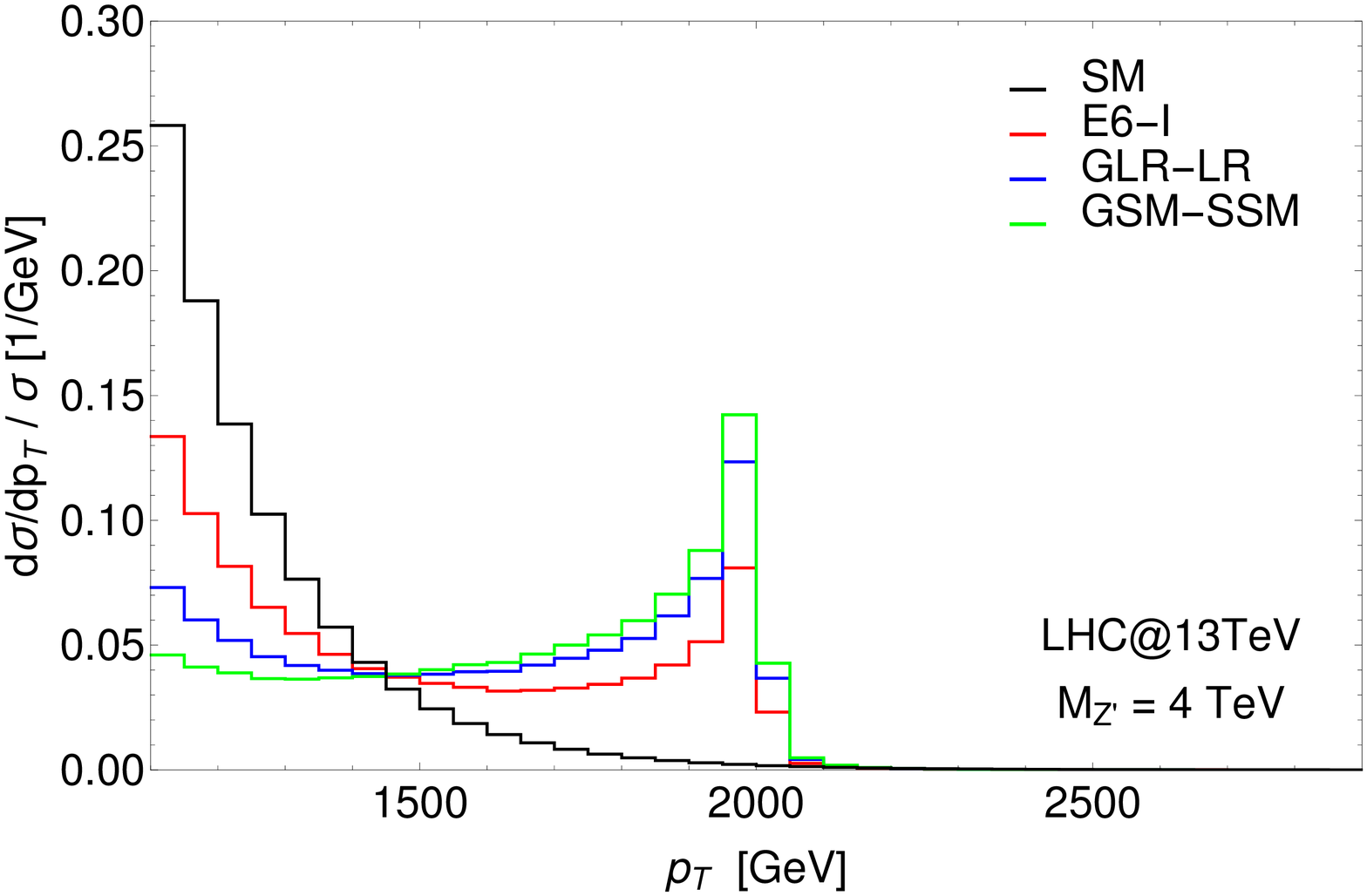}{(b)}
\includegraphics[width=0.30\textwidth]{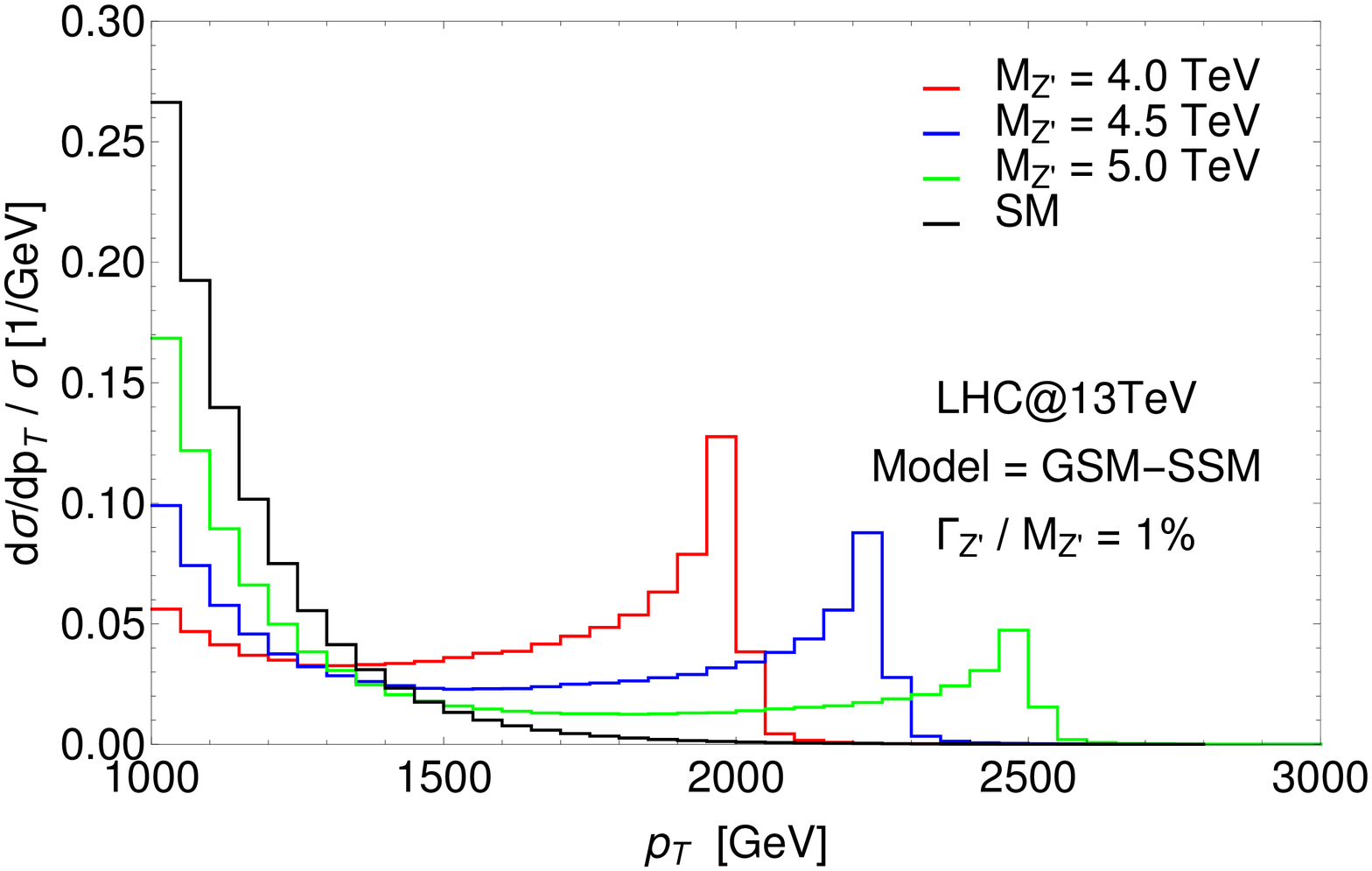}{(c)}
\caption{Normalised $p_T$ distribution of either lepton as predicted in the SM and in the three benchmarks fixing $M_{Z^\prime}$ = 4 TeV and
$\Gamma_{Z^\prime} / M_{Z^\prime}$ = 1\%, in the same setup as before but changing 
(a) the LHC collider energy to 8 TeV,
(b) the $p_T^{\rm min}$ = 1100 GeV,
and (c) in the SSM fixing the $Z^\prime$ mass to 4, 4.5 and 5 TeV.}
\label{fig:FP_dependence}
\end{center}
\end{figure}

The position of the FP indeed depends only on few parameters, that are summarised in Fig.~\ref{fig:FP_dependence}.
In Fig.~\ref{fig:FP_dependence}(a) we changed the collider energy to 8 TeV and the FP consequently changed.
Anyway from now on we will consider only the setup of the LHC at 13 TeV, as this is and will be the environment of current and future analysis.
It has been verified~\cite{Accomando:2017fmb} that the effects of kinematical cut (such as $|\eta|$ acceptance cuts) as well as interference effects 
are totally irrelevant in our analysis and they do not affect the appearance of the FP, neither they spoil any of its features.

Fixing the collider centre-of-mass energy, we have noticed that the position of the FP is then determined only from two parameters which are the 
choice of the $p_T^{\rm min}$ and the mass of the $Z^\prime$.
An empirical relation has been derived from the results of the simulations:

\begin{equation}
 FP = p_T^{\rm min} + 10\%M_{Z^{\prime}}.
 \label{eq:FP}
\end{equation}

Since the FP arises from the normalisation procedure defined above, it is by all means determined by the choice of the $p_T^{\rm min}$, 
assuming the above condition on the $p_T^{\rm max}$.
In Fig.~\ref{fig:FP_dependence}(b) the lower limit on the normalisation has been fixed to $p_T^{\rm min}$ = 1100 GeV and the position of the FP
has scaled accordingly to Eq.~\ref{eq:FP}.
The other relevant parameter to the FP position is the $Z^\prime$ mass, that we have varied in Fig.~\ref{fig:FP_dependence}(c) between 4 and 5 TeV,
which is the $M_{Z^\prime}$ region that will be covered in ongoing analysis.
Again the position of the FP, that now can be identified by the crossing point between the SM and BSM curves, follows the empirical relation
derived in Eq.~\ref{eq:FP}.

\section{The Asymmetry of the Focus Point}

\begin{figure}[h]
\begin{center}
\includegraphics[width=0.45\textwidth]{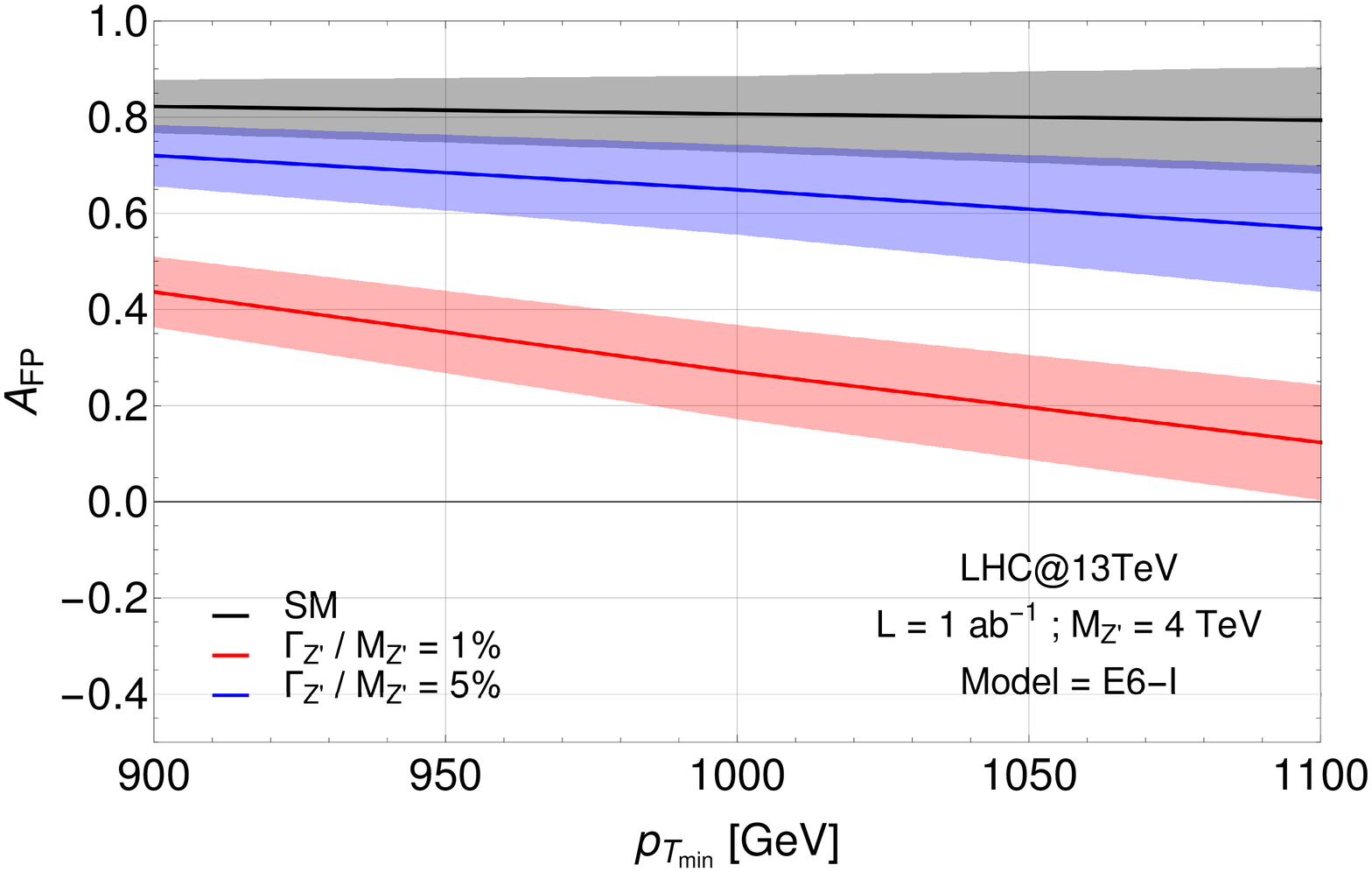}{(a)}
\includegraphics[width=0.45\textwidth]{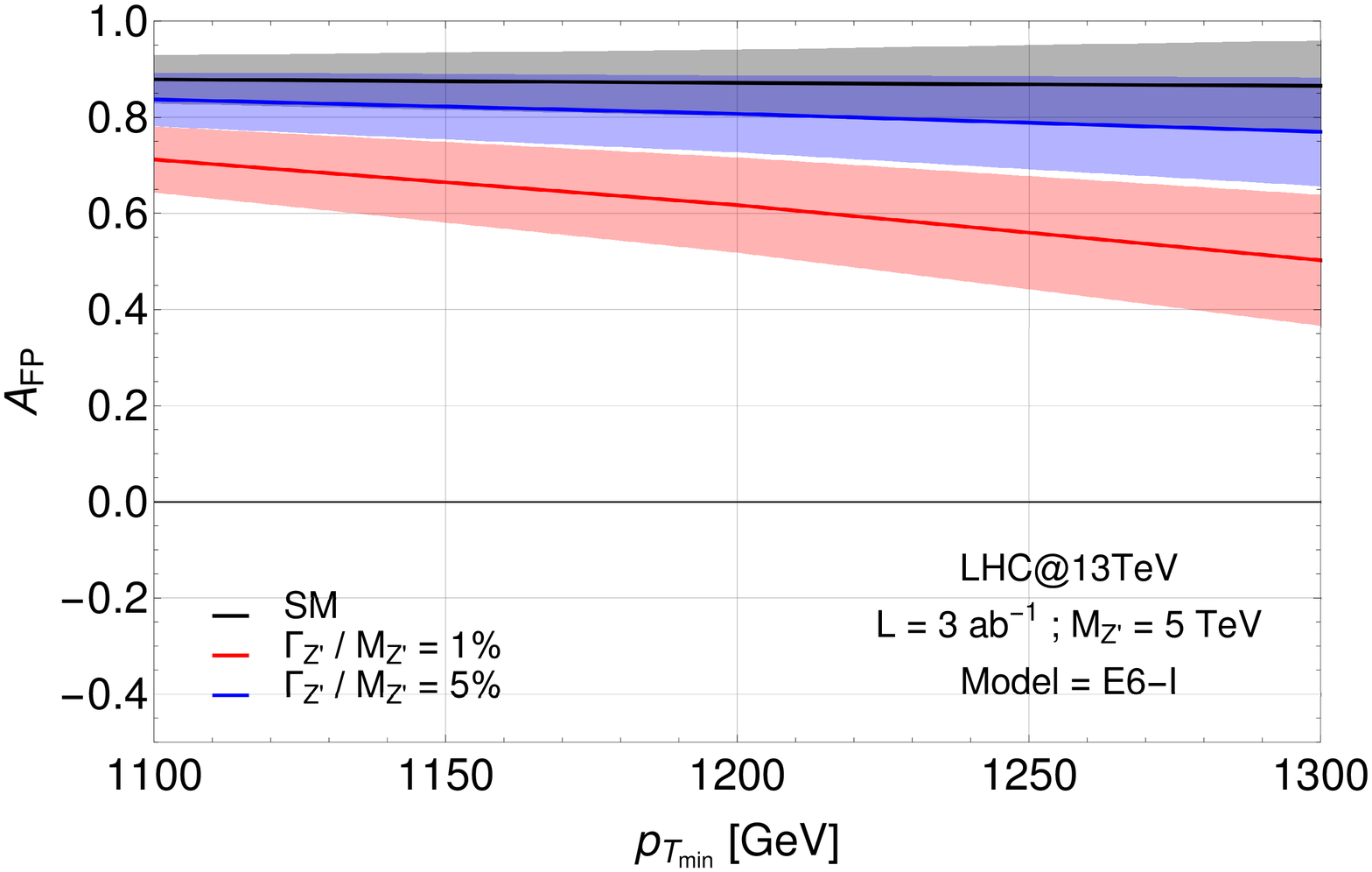}{(b)}
\includegraphics[width=0.45\textwidth]{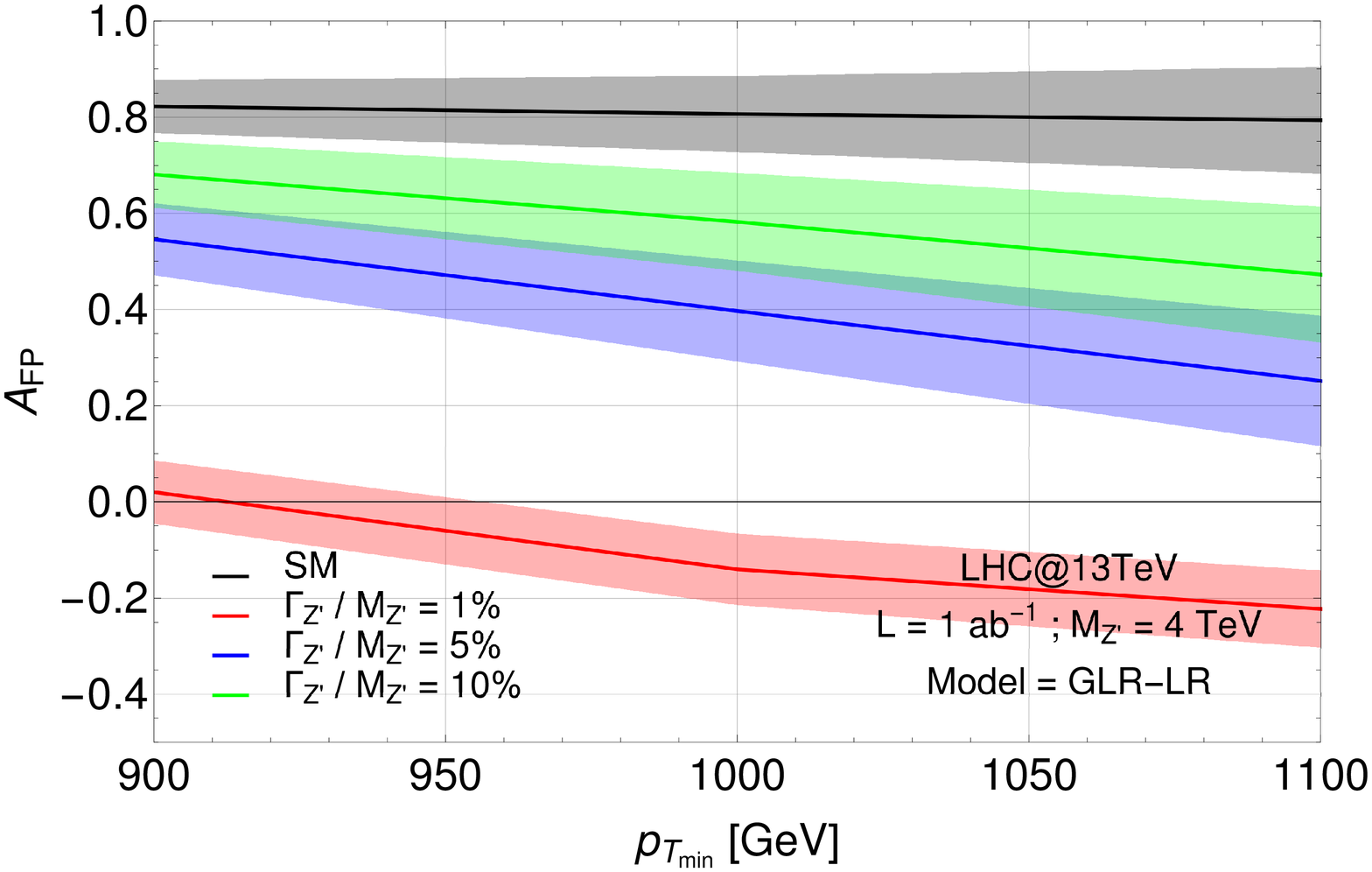}{(c)}
\includegraphics[width=0.45\textwidth]{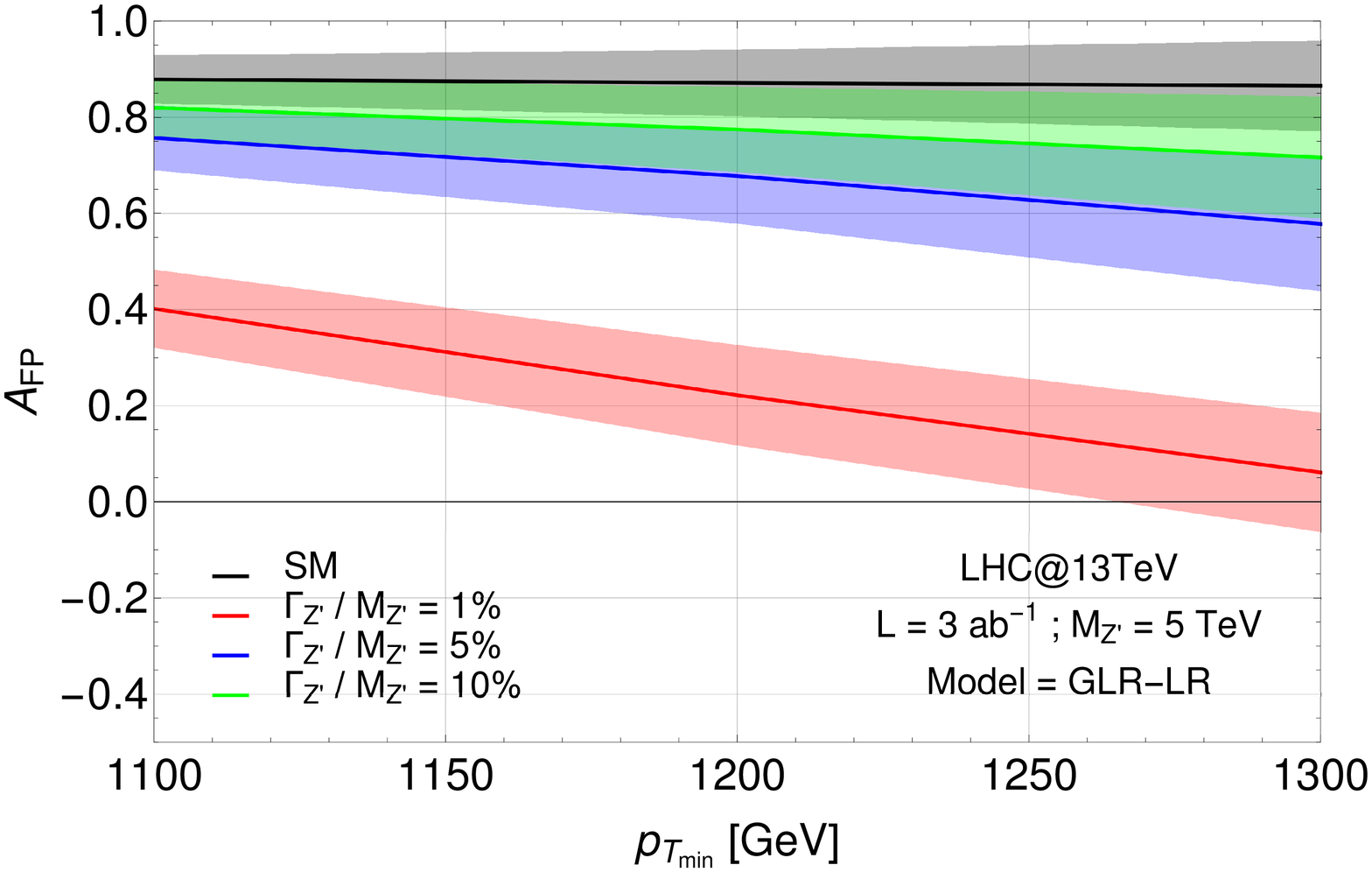}{(d)}
\includegraphics[width=0.45\textwidth]{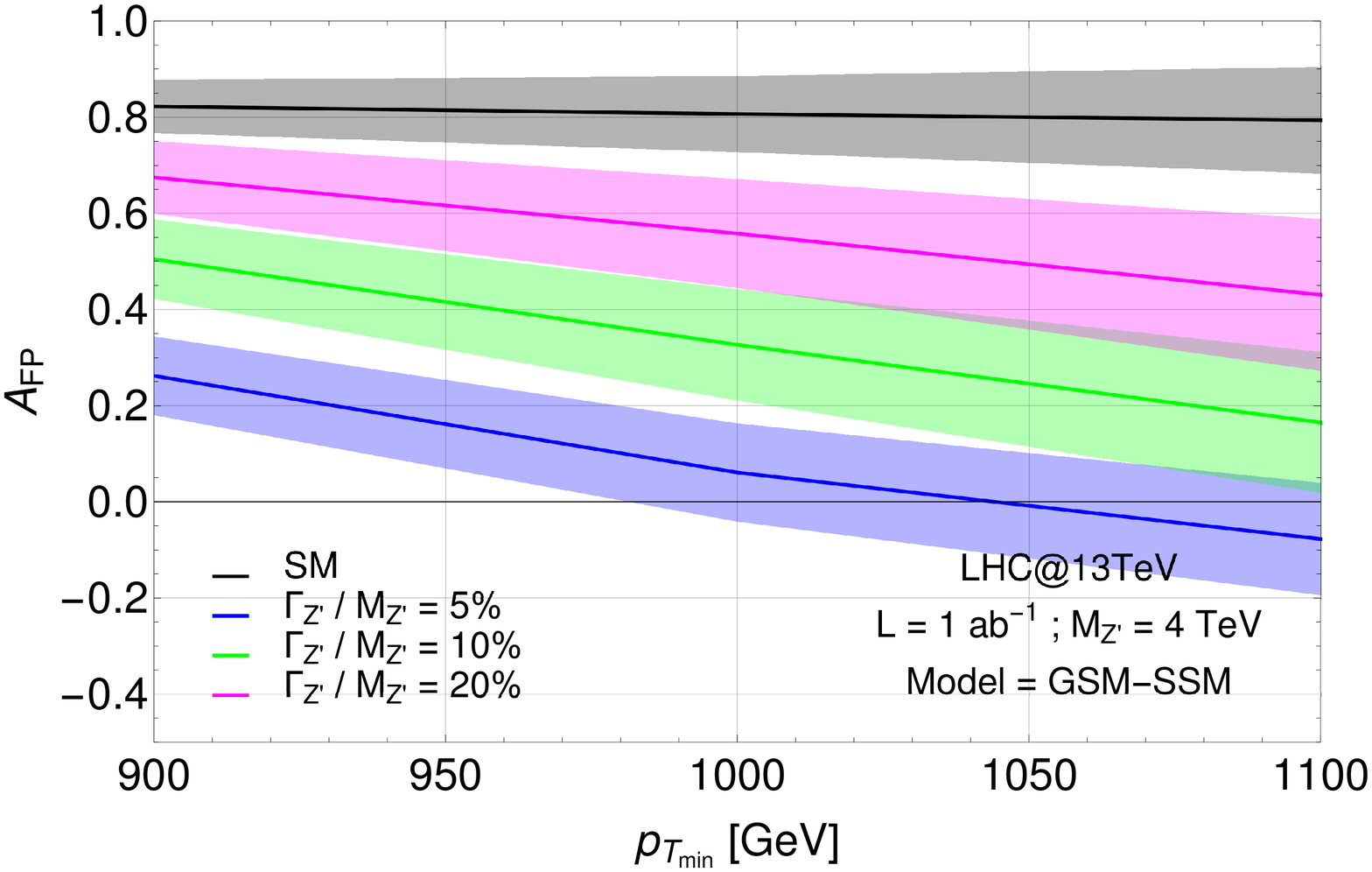}{(e)}
\includegraphics[width=0.45\textwidth]{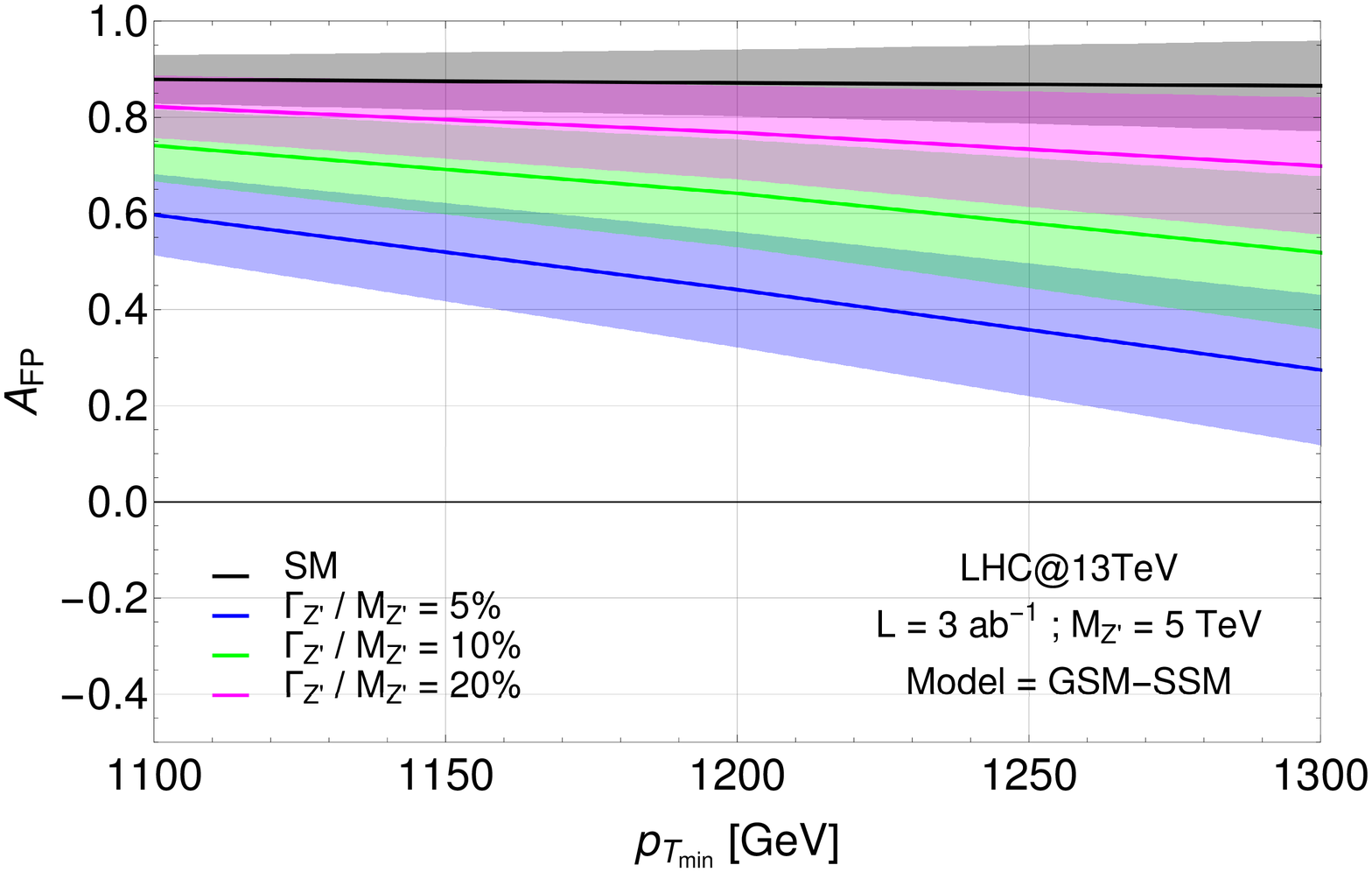}{(f)}
\caption{$A_{\rm FP}$ central value and statistical 1$\sigma$ error band as function of $p_T^{\rm min}$ cut for the LHC at 13 TeV 
and $\mathcal{L}$ = 1 {\rm ab}$^{-1}$. The black line represents the SM while the coloured lines represent four different widths (1\%, 5\%, 10\% 
and 20\%) of the $Z^\prime$ resonance in the (a) {\rm E6-I}, (c) {\rm GLR-LR} and (e) {\rm GSM-SSM} model fixing $M_{Z^\prime}$ = 4 TeV.
Same exercise is repeated in (b), (d) and (f) fixing $M_{Z^\prime}$ = 5 TeV and $\mathcal{L}$ = 3 {\rm ab}$^{-1}$.
The values of the FP are chosen in accordance to Eq.~\ref{eq:FP}.
}
\label{fig:AFP_widths}
\end{center}
\end{figure}

Now consider the interval in the $p_T$ spectrum that has been chosen for the above normalisation.
The FP can be used to separate the interval in two region, on the ``left" and on the ``right" side of the FP.
We define the $L$ and $R$ quantities as the integrals of the curves of Fig.~\ref{fig:pT_norm} in the two corresponding subregions:

\begin{equation}
L=\frac{1}{N}\int_{p_T^{\rm min}}^{\rm FP}\frac{d\sigma}{dp_T}dp_T,\quad\quad
R=\frac{1}{N}\int_{\rm FP}^{p_T^{\rm max}}\frac{d\sigma}{dp_T}dp_T.
\end{equation}

The new observable that will be called Asymmetry of the Focus Point (or $A_{\rm FP}$) is defined as the normalised difference between
these two integrated quantities:

\begin{equation}
 A_{\rm FP}=\frac{L-R}{L+R}.
\end{equation}

For a given resonance (\ie~a $Z^\prime$ with a fixed mass and width) the $A_{\rm FP}$ only depends on the choice of the $p_T^{\rm min}$ by construction.
In the situation where a $Z^\prime$ signal is observed, the new observable can be used to support the evidence and its measurement provides an
independent information on the underlying model generating the new boson and more importantly on the width of the resonance.

In order to test the sensitivity of the $A_{\rm FP}$, in each plot of Fig.~\ref{fig:AFP_widths} we are showing 
its value and its statistical error for different choices of the resonance width, as function of the choice of the $p_T^{\rm min}$
(the value that has been used for the position of the FP follows Eq.~\ref{eq:FP}).
This is repeated for three benchmark models ({\rm E6-I}, {\rm LR} and {\rm SSM}), each time for for two different values of the $Z^\prime$ mass
and integrated luminosity (4 TeV with $\mathcal{L}$ = 1 {\rm ab}$^{-1}$ and 5 TeV with $\mathcal{L}$ = 3 {\rm ab}$^{-1}$).
As visible comparing Fig.~\ref{fig:AFP_widths}(a), (c) and (e), or the corresponding plots with $M_{Z^\prime}$ = 5 TeV,
if we assume to know the mass of the resonance (easily inferable from the invariant mass distribution) and its width, 
a measurement of the $A_{\rm FP}$ can be useful to disentangle the underlying BSM model.
However the three benchmark models chosen here are representative of their respective classes
({\rm E6}, Generalised Left-Right or {\rm GLR} and Generalised Standard Model or {\rm GSM}), in the sense that
it has been verified~\cite{Accomando:2017fmb} that the value of the $A_{\rm FP}$ is similar for each model within the same class,
such that the observable is able to distinguish between different classes of models, but has almost no sensitivity in distinguishing the
particular model within each class.

More importantly is the case where we fix the BSM construction and the $Z^\prime$ mass, while we seek for information on the resonance width.
This would be a common scenario imagining that a $Z^\prime$ peak is observed and various benchmark models would be used to fit the signal.
In this framework the $A_{\rm FP}$ can be used to derive important constrains on the resonance width, which can be imported as independent 
constrain in the experimental fit of the BW shape in the invariant mass spectrum.
As visible in Fig.~\ref{fig:AFP_widths} the $A_{\rm FP}$ can constrain resonances widths up to $\Gamma_{Z^\prime} / M_{Z^\prime}$ = 5\%,
$\Gamma_{Z^\prime} / M_{Z^\prime}$ = 10\% and $\Gamma_{Z^\prime} / M_{Z^\prime}$ = 20\% respectively within the {\rm E6}, {\rm GLR} and {\rm GSM}
classes of models.

\section{Conclusions}

We have introduced a new observable, the $A_{\rm FP}$ that can be used for $Z^\prime$ diagnostic in the di-lepton final channel.
The definition of the novel observable goes through simple measurements on the transverse momentum spectrum of either lepton in the final state.

We have shown that a FP appears once normalising the $p_T$ distributions obtained for various $Z^\prime$ benchmark models with different widths
and for the SM.
The position of the FP indeed does not depend on the underlying BSM construction neither on the resonance width.
By construction it only depends on the $Z^\prime$ mass and on the $p_T^{\rm min}$, which defines the interval chosen for the normalisation
(once we assume $p_T^{\rm max} > M_{Z^\prime}/2$).
We have derived an empirical relation between the position of the FP, the mass of the $Z^\prime$ and the chosen $p_T^{\rm min}$,
which holds for the LHC with 13 TeV c.o.m. energy and for the $M_{Z^\prime}$ parameter region of concern during the Run-II and the HL-LHC stage.

Exploiting the FP and its properties, we have defined the $A_{\rm FP}$ observable as the normalised difference between two (normalised) differential
cross sections integrated in two regions of the $p_T$ spectrum separated by the FP.
As already said, the position of the latter is both model and width independent, thus assuming to have a fixed collider energy and the value of the 
$Z^\prime$ mass from invariant mass measurements, the $A_{\rm FP}$ observable is well defined and its measured value will depend only on the choice
of the $p_T^{\rm min}$ and on the resonance width.

Being an integrated observable, the $A_{\rm FP}$ exploits the full data set and its statistical uncertainty is therefore small,
and consequently it also does not require any modelling of the shape.
Moreover, being a ratio of cross sections, part of the systematic uncertainties affecting the $A_{\rm FP}$ naturally cancel.
In this sense it can be useful to support the diagnostic analysis of an excess of events observed in the di-lepton high invariant mass, as it provides 
independent constrains on the $Z^\prime$ width, that can improve the quality of experimental fits.
The properties we have described for the $A_{\rm FP}$ make it a very attractive search tool as well, as it still maintain a good sensitivity on 
new physics also in the problematic context of broad resonance detection.

\end{document}